\documentstyle[twocolumn,aps,prl,epsf,floats]{revtex}
\newcommand{\la}[1]{\label{#1}}

\newcommand{\be}{\begin{equation}}
\newcommand{\ee}{\end{equation}}
\newcommand{\bi}{\begin{itemize}}
\newcommand{\ei}{\end{itemize}}
\newcommand{\ba}{\begin{eqnarray}}
\newcommand{\ea}{\end{eqnarray}}
\newcommand{\bea}{\begin{eqnarray}}
\newcommand{\eea}{\begin{eqnarray}}

\newcommand{\roots}{\sqrt{s}}

\newcommand{\qsatk}{Q_{S,A}}
\newcommand{\qsaty}{Q_{S,p}}

\def\lsim{\,\raise0.3ex\hbox{$<$\kern-0.75em\raise-1.1ex\hbox{$\sim$}}\,}
\def\gsim{\,\raise0.3ex\hbox{$>$\kern-0.75em\raise-1.1ex\hbox{$\sim$}}\,}

\begin{document}
\twocolumn[\hsize\textwidth\columnwidth\hsize\csname
@twocolumnfalse\endcsname
 
\draft
\title{Saturation and Pion Production in Proton-Nucleus Collisions}

\author{
 J.T. Lenaghan$^{a}$ and
 K. Tuominen$^{b}$
}
\address{$^{a}$ The Niels Bohr Institute and $^{b}$ NORDITA, \\} 
\address{Blegdamsvej 17, DK-2100 Copenhagen \O, Denmark} 
\date{August 1, 2002}
\maketitle
 
\vspace*{-3.5cm}
\noindent
\hfill
\mbox{NORDITA-2002-35 HE, hep-ph/0208007}
\vspace*{3.5cm}

\begin{abstract}\noindent
We study the effects of gluon saturation on pion production in high
energy proton-nucleus collisions using the color glass condensate
model.  At high $p_\perp$, we show that the $p_\perp$-distribution of
gluons behaves as $\sim 1/p_\perp^7$ in accordance with both
conventional perturbative QCD calculations and experiment.
Fragmentation of gluons into pions leads to a rapidity dependent
depletion of pions relative to the conventional perturbative QCD
predictions. We argue that these clear and systematic differences
provide a signal for the onset of gluon saturation which is
accessible in upcoming experiments.
\end{abstract}

\vspace*{-0.1cm}

\vskip1.5pc]

\vspace*{-0.5cm}

\narrowtext

Since the pioneering work of Ref.\ \cite{glr}, the phenomenon of gluon
saturation, occurring at very high gluon density, has been under
intense study.  One hopes that saturation would naturally tame the rise
of the gluon distribution function at small values of Bjorken $x$ and
circumvent the subsequent violation of unitarity.

A promising attempt to address the problem of saturation involves
replacing the partonic fields by coherent classical fields. The
justification is that for large occupation numbers the dynamics are
essentially classical. This program was initiated in Ref.\
\cite{McLerran:1993ni} and developed further in e.g.  Refs.\
\cite{Kovchegov:1998bi,McLerran:1994vd}.  Within this approach one is
able to show that an intrinsic scale, the saturation scale
$Q_{S,A}(x)$, is generated. This scale depends on both Bjorken $x$ and
atomic number, $A$, and roughly delineates the dense and dilute
regimes. Throughout this work we shall refer to this approach as the
color glass condensate (CGC) framework.

The parametric dependence of the saturation scale has been extensively
studied. Phenomenological models based on the idea of parton
saturation have been successful in describing both the inclusive and
diffractive HERA data for $F_2$ \cite{Golec-Biernat:1999qd}. According
to this model, the dependence of the saturation scale on Bjorken $x$
is \be
\label{eq:satscale} \qsatk^2(x) = (1 {\rm GeV} )^2
\left(\frac{x_0}{x}\right)^\lambda \, A^{1/3}, \ee where $\lambda =
0.25\dots 0.3$ and $x_0=3\cdot 10^{-4}$. For the numerical
calculations, we will use $\lambda=0.3$. Since $\qsatk^2$ can be
interpreted as the transverse density of partons, it is enhanced by
considering larger nuclei at any given $x$ by a factor of $A^{1/3}$.
As a result, one may expect to observe saturation effects better in
nuclear collisions at currently available collider energies. Indeed,
the present data from BNL--RHIC
\cite{multiPHENIX,multiSTAR,multiPHOBOS,multiBRAHMS} seem to be
compatible with saturation based phenomenology
\cite{Eskola:1999fc,Kharzeev:2000ph,KL,Larrylect}. These results,
however, are far from being conclusive on account of numerous model
dependencies related to our poor knowledge of the details of the
spacetime evolution of the produced strongly interacting matter.

The onset of saturation has important qualitative and quantitative
consequences for the $p_\perp$-distribution of the produced
gluons. This was already pointed out in Ref.\ \cite{glr} in the
context of proton--proton collisions where an asymmetry in the
saturation scales of the two protons appears away from the
mid-rapidity.  As one hadron moves deeper into the saturation region,
towards smaller $x$, its saturation scale increases, while the
saturation scale of the second hadron decreases as is evident from
Eq.\ (\ref{eq:satscale}).  In proton--nucleus collisions, a similar
situation is realized already at mid-rapidity due to the atomic number
asymmetry.  In particular, one expects three different regions in the
transverse momentum distribution of the produced gluons on account of
the asymmetries in the saturation momenta, $\qsatk > \qsaty$.  There
is the region $p_\perp < \qsaty < \qsatk$ in which both the proton and
the nucleus are saturated.  The second possibility is that the nucleus
is saturated but the proton is in the linear regime ($\qsaty < p_\perp
< \qsatk$).  The final region is the usual perturbative domain in
which both gluon distributions are entirely governed by the linear
evolution ($\qsaty < \qsatk < p_\perp $).  Consequently,
proton--nucleus collisions offer the best possibility of
experimentally probing the systematics of gluon saturation. The
problem of evaluating gluon production in proton--nucleus collisions
has been found to be analytically tractable above the smaller
saturation scale and in Ref.\ \cite{dumiMcL} this was qualitatively
studied in the CGC framework. 

In this Letter, we undertake the quantitative computation of the
produced hadronic spectra based on the framework of Ref.\
\cite{dumiMcL} and compare it to the conventional perturbative QCD
(pQCD) calculation which does not include any saturation effects.  The
main emphasis is on a numerical calculation of the fragmentation of
the gluon distributions into pions, and on the study of how saturation
effects manifest themselves in the $p_\perp$-distributions of the
measured hadrons.  We shall see that although the spectra at the
gluonic level are qualitatively very different, this difference is
made less striking by the fragmentation of the gluons into pions.

In pQCD, the inclusive cross section for gluon production is given as
a convolution of the hard partonic scattering cross section with the
gluon distribution functions.  The hard cross section behaves as $\sim
1/p_\perp^4$, but the final result receives additional contributions
from the two gluon distributions and from the running of the strong
coupling constant.  As a result, the slope of the transverse momentum
distribution actually falls much more rapidly with a scaling closer to
$\sim 1/p_\perp^7$.  Any realistic model of particle production must
reproduce this scaling at large transverse momenta.

We consider a CGC model obtained by combining two separate limits of 
the results of Ref.\ \cite{dumiMcL}. The gluon production cross section
in the perturbative domain is given by
\be 
\frac{d\sigma^{\rm{pert}}_{g}}{d^2p_\perp dy}=
\frac{8 N_c(N_c^2-1)}{\pi}\frac{\alpha_s^3}{p_\perp^4}\chi_p(x_{p},p_\perp^2)
\chi_A(x_{A},p_\perp^2),
\label{pertxsect} 
\ee 
where 
\ba \nonumber
\chi_i(x_{i},p_\perp^2) &=& \frac{1}{\pi R_{i}}\frac{N_c}{N_c^2-1}\\ \nonumber
&\times&\left(\int_{x_{i}}^1dx
g_{i}(x,p_\perp^2)+\frac{C_F}{N_c} \int_{x_{i}}^1dx
q_{i}(x,p_\perp^2)\right), 
\ea 
with $i=A,p$. $R_i$ is the transverse radius,
$x_{A,p}=p_\perp e^{\mp y}/\roots$, $C_F=(N_c^2-1)/(2N_c)$ and the 
nucleus is chosen to have negative beam rapidity.  
Here $q_{i}$ and $g_{i}$ denote
the valence quark and gluon distributions in the nucleus or in the
proton. One needs to evaluate the color charge functions $\chi_i$
only in the weak field regime, and therefore the use of DGLAP evolved
parameterizations is appropriate.

In the saturation domain of the 
nucleus, the cross section is given by
\be
\frac{d\sigma^{\rm{sat}}_{g}}{d^2p_\perp dy}=
\frac{C(N_c^2-1)}{2\pi^2}\frac{\alpha_s}{p_\perp^2}\chi_p(x_{p},p_\perp^2)
\pi R_A^2, 
\label{satxsect}
\ee 
The constant $C$ 
is chosen so that the sub-cross section is continuous at $\qsatk$. We
have neglected the logarithms appearing in the results of
Ref.\ \cite{dumiMcL} since one expects the overall behavior 
to be driven by the strong powers of $p_\perp$ and the constant under the
logarithm is in any case difficult to fix. The overall normalization 
is unimportant as we will discuss later. 

From these equations, it is clear that retaining the DGLAP evolved
parton distributions for the weak fields and
letting the strong coupling run will affect the qualitative results
of Ref.\ \cite{dumiMcL} which display the expected $\sim 1/p_\perp^4$
and $\sim 1/p_\perp^2$ behavior in the perturbative domain and in the
saturation domain of the nucleus, respectively. The inclusion of these
effects is essential for making the proper contact with conventional
result for high $p_\perp$ mentioned above. 

In order to compute the produced hadronic spectra, we convolute the
above inclusive gluon production cross section with the appropriate
fragmentation functions. For the sake of simplicity, we do not attempt
to account for the Cronin effect \cite{Cronin:zm}, nuclear shadowing
\cite{Arneodo:1992wf} or nuclear modifications of the fragmentation
functions \cite{Guo:2000nz}. So, $xg_A(x,Q^2)=Axg_p(x,Q^2)$
in the above equations
and similarly for quarks.  The
generalization to nucleus--nucleus collisions with $A_1\ll A_2$ is
trivial through replacements $p\rightarrow A_1$, $A\rightarrow
A_2$. We note that the overall normalization of the results 
is difficult to fix. In
Ref. \cite{EH} it was found that in order to reproduce the data from
hadronic collisions above a cutoff momentum of $p_0\sim 1\dots 3$ GeV
using pQCD one needs a $\roots$-dependent $K$-factor to account for
the higher order corrections. The relative normalization between the
pQCD and CGC results, however, is almost fixed since both results
should match at high $p_\perp$. Therefore, we do not attempt to
fine-tune the overall normalization, but rather consider a leading order pQCD
calculation, including only gluons, without any $K$-factors and point
out the differences relative to the CGC calculation which will be
normalized at high $p_\perp$ 
to the pQCD result. Our central results, the slopes of the
$p_\perp$- and $y$-distributions, are not sensitive to the
normalization ambiguities. One should note, however, that the $\roots$
growth of the gluon multiplicity in the CGC model is driven by the
small $x$-growth of the gluon distributions, and is therefore of the
order of $dN/dy\sim\roots^\lambda$. We will focus exclusively on pions
since this distribution approximates well the distribution of all
hadrons.

To compute the fragmentation of gluons into pions, 
we assume that the pion is produced collinearly with its parent gluon, 
$\eta_g=\eta_\pi$, and carries a fraction $z$ 
of the parent gluon's energy. 
The cross section for the production of pions is
\be
\frac{d\sigma^{p A \rightarrow \pi X}}{d^2q_\perp dy}=
J(m_\perp,y)\int\frac{dz}{z^2}D_g^\pi(z,q_\perp^2)
\frac{d\sigma^{p A\rightarrow g X}}{d^2p_\perp dy_g}, 
\ee 
where 
\ba
\nonumber  & & p_\perp=\frac{q_\perp}{z}J(m_\perp,y),~
y_g = \sinh^{-1}\left(\frac{m_\perp}{q_\perp}\sinh y\right),\\ \nonumber
& & J(m_\perp,y) = \left(1-\frac{m_\pi^2}{m_\perp^2\cosh ^2y}\right)^{-1/2} 
\ea 
and $m_\perp^2=q_\perp^2+m_\pi^2$.
The integration over $z$ is limited by the
maximal energy the gluon can carry and by the lower bound 
on $p_{\perp}$. In pQCD, this latter scale
is a fixed cutoff, while in the CGC calculation it is  
determined by the saturation scale of the proton, 
\be 
\frac{a m_\perp}{\roots}\cosh y\leq z\leq\min\left(1,\frac{q_\perp}
{p_{\perp,min}}J(m_\perp,y)\right) \,\, ,
\ee 
where $a=1,2$ for the 
CGC and pQCD calculations respectively. This difference
arises from the underlying parton kinematics.  
In LO pQCD, one produces
two minijets, back-to-back in the transverse plane, carrying at most
$E=\roots/2$. In the CGC calculation, the scattering is a
BFKL-type $2\rightarrow 1$ fusion, and the produced gluon can have
energy up to $E=\roots$.
We use the CTEQ5 parameterization of
parton distributions \cite{CTEQ5} and KKP parameterization of the
fragmentation functions \cite{KKP}. 

In Fig.\ \ref{ptgluons}, we plot the $p_\perp$-distributions of the
produced gluons at different rapidities. The topmost three curves are
for $y=0$ and the lower two are for $y=3$. The dashed curve is the CGC
result including only gluons in the sources $\chi_i$.  Note that at
$y=0$ the slopes of the CGC and pQCD calculations almost match at high
$p_\perp$. Below the saturation scale of the nucleus, the two results
strongly deviate. In a full computation of the gluon distribution,
this sharp bend at $Q_{S,A}(y)$ will become smooth. The essential
point is that there is a marked depletion of gluons for $p_\perp <
Q_{S,A}(y)$.

\begin{figure}[ht]

\vspace{-2.5cm}

\epsfysize=8.5cm
\centerline{\epsffile{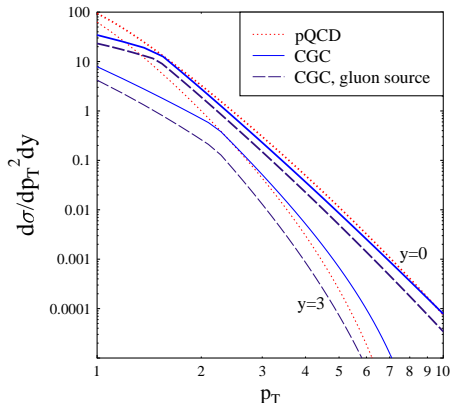}}

\vspace{-0.8cm}

\caption[a]{Transverse momentum distributions of gluons produced in a
proton--gold collision computed from conventional pQCD (dotted) and
from the CGC calculation (solid).  The uppermost set of curves is for
$y=0$, while the lower ones correspond to $y=3$. The dashed curve shows the
CGC result with purely gluonic source.}

\vspace{-0.0cm}
\la{ptgluons}
\end{figure}

\begin{figure}[ht]

\vspace{-2.5cm}

\epsfysize=8.5cm
\centerline{\epsffile{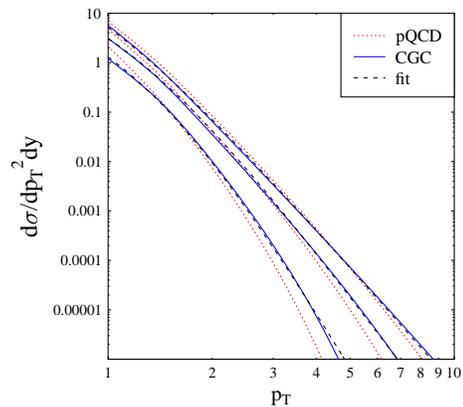}}

\vspace{-0.8cm}

\caption[a]{Transverse momentum distributions of pions produced in a
proton--gold collision computed by fragmenting gluons from
conventional pQCD (dotted) and from CGC (solid) calculations.  The
topmost two curves are for $y=0$ and the other pairs are for $y=2$ and
$y=3$, respectively.  The dashed line shows the suggested fit of the
pion distribution.}

\la{ptpions}
\end{figure}

If this spectrum were measured, one should be able to determine whether
the parton dynamics in the collision are better described by pQCD or
by the CGC model.  To obtain a measurable spectum one needs to
fragment each of the distributions in Fig.\ \ref{ptgluons} to hadrons.

In Fig.\ \ref{ptpions}, we show the $p_\perp$-distribution of the
pions produced from the two models.  The fragmentation changes the
qualitative behavior of the spectrum of the CGC model dramatically
with the sharp bend disappearing completely on account of the
redistribution of momentum from the gluons to the pions.  We note that
there are significantly fewer produced pions at lower transverse
momentum in the CGC calculation than in the conventional quantum pQCD
one and that this deficit is strongly rapidity dependent.  We have
checked using a smooth fit interpolating between the two limiting
forms of the gluon distribution that the qualitative, and to some
extent even quantitative, results for the pion spectrum in the CGC
model are not sensitive to the sharpness of the knee in the gluon
distribution.

To obtain an useful simple parameterization of the pion spectrum, we
consider the form \be
\frac{d\sigma}{dydp_\perp^2}=\frac{C(y)}{(p_\perp^2+\mu(y)^2)^{b(y)}},
\label{eq:pionfit}
\ee
where the emergence of a rapidity dependent scale, $\mu(y)$, is a natural
consequence of the rapidity dependence of the saturation scale. The
rapidity dependence of the effective slope $b(y)$ is expected since
the slope of the gluon distribution already depends on rapidity.
We find that $\mu(y)^2=0.50e^{\lambda y}$, where $\lambda=0.3$. The
values of the fit parameters $C(y)$ and $b(y)$ at various rapidities
are collected in Table\ \ref{fittable}. The inclusion of the quarks plays
a role for the overall normalization and the slopes are left almost unchanged 
as can be seen already from Fig. \ref{ptgluons}

\begin{table}
\vspace*{-0.5cm}
\center
\begin{tabular}{p{1cm} p{1cm} p{1cm}}
\hline
$y$    & $C(y)$  & $b(y)$ \\
\hline
0    & 30.0  & 4.00 \\
0.5  & 40.0  & 4.10 \\
1.0  & 45.0  & 4.20 \\
1.5  & 60.0  & 4.40 \\
2.0  & 80.0  & 4.70 \\
2.5  & 130.0 & 5.30 \\
3.0  & 220.0 & 6.00 \\
\hline 
\end{tabular}
\caption[1]{\protect\small The fit parameters $C(y)$ and $b(y)$ in 
Eq.\ (\ref{eq:pionfit}) at different rapidities.}
\vspace*{-0.5cm}
\la{fittable}
\end{table}


\begin{figure}[ht]

\vspace{-2.2cm}

\epsfysize=8.5cm
\centerline{\epsffile{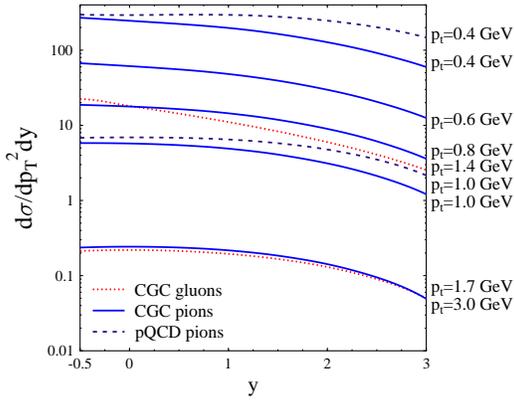}}

\vspace{-0.8cm}

\caption[a]{The rapidity densities at fixed $p_\perp$. The dotted curves
show the spectrum of gluons in the CGC model at $p_\perp=1.4$ GeV
(upper) and $p_\perp=3$ GeV (lower). The solid curves give the pion spectra
from CGC model at $p_\perp=0.4, 0.6, 0.8, 1.0, 1.7$ GeV (top to
bottom), and the dashed ones show the pion spectra from pQCD at
$p_\perp=0.4$ (upper) and 1.0 GeV (lower).} 
\vspace{-0.5cm}
\la{fig:rap}
\end{figure}
Finally, we consider the rapidity distributions at fixed
$p_\perp$. One cannot approach the beam rapidity of the nucleus 
in this framework, since at such large
negative rapidities the saturation scales of the nucleus and the proton
become comparable and the underlying model breaks down. Near
midrapidity, the CGC model gluon spectrum in perturbative domain is
characterized by constant behavior in rapidity, as is the case for
pQCD spectrum. In the saturation domain, the CGC gluon spectrum 
has a nonzero slope also at midrapidity. The CGC model gluon spectra
are shown in Fig. \ref{fig:rap} with dotted curves for $p_\perp=3$ and
1.4 GeV, and the corresponding pion spectrum by solid lines for
$p_\perp=0.4,~0.6~0.8~1.0$ and 1.7 GeV. The pion spectrum from pQCD,
shown by dashed lines for 0.4 and 1.0 GeV has the flat behavior at
midrapidity over the whole $p_\perp$ range. We observe that the small
$p_\perp$ pions might reveal the characteristic form carried by the
gluon spectrum in the saturation domain. This result is easy to
understand, since the pion spectrum at scale $q_\perp$ effectively
probes the gluon spectrum at some higher scale $q_\perp/\bar{z}$. For
the perturbative distribution one finds that $\bar{z}\sim 0.5\dots
0.6$. This is clearly seen by comparing pions at $p_\perp=1.7$ GeV and
gluons at $p_\perp=3$ GeV. As the slope of the gluon distribution
decreases, $\bar{z}$ moves towards smaller values as well. This causes
the gluons in saturation domain to fragment mainly to much smaller
values of $p_\perp$ than the perturbative gluons would. The situation
improves at LHC energies where the saturation domain extends already
at midrapidity to $\sim 2$ GeV. At large rapidities it is difficult to
distinguish between saturation and perturbative spectra, and
presumably different dynamics which are not contained in either of
these models become dominant towards the fragmentation regions.

We have investigated the quantitative behavior of the $p_\perp$- and
$y$-distributions of gluons and pions in collisions of nuclei with a
large atomic number asymmetry. We have shown that within the color
glass condensate framework one indeed reproduces the behavior of the
conventional pQCD at high $p_\perp$ which significantly deviates from
commonly cited expectation of $1/p_\perp^4$ in the literature on
saturation. The fragmentation of gluons into pions was shown to
significantly change the qualitative shape of the spectrum. However, a
clear difference from the spectrum obtained from the conventional pQCD
calculation in the form of a rapidity dependent pion deficit can still
be extracted by measuring the rapidity dependence of the
$p_\perp$-distributions. The rapidity dependence of the slopes were
found to be different at small $p_\perp$, and a change in the form of
the rapidity distribution of pions as a function of $p_\perp$ could be
used to determine the onset of saturation.

{\bf{Acknowledgements:}} 
We would like to thank A.\ Dumitru, K.J.\ Eskola,
K.\ Kajantie, J.\ Jalilian--Marian, L.\ McLerran and W.\ Vogelsang for useful 
discussions.  J.T.L.\ thanks the Nuclear Theory Group 
at Brookhaven National Laboratory, where part 
of this work was done, for hospitality.

\end{document}